\documentclass[twocolumn,showpacs,preprintnumbers,pre,floatfix,superscriptaddress,amsmath,amssymb]{revtex4}
\usepackage{epsfig}
\usepackage{subfigure}
\usepackage{amsmath}
\usepackage{amssymb}
\usepackage{graphicx}
\usepackage{dcolumn}
\usepackage{bm}
\input epsf

\begin{document}

\title{Approximating the largest eigenvalue of network adjacency matrices}

\author{Juan G. Restrepo}
\email{juanga@neu.edu} \affiliation{Center for Interdisciplinary
Research in Complex Systems, Northeastern University, Boston, MA
02115, USA }

\author{Edward Ott}
\affiliation{ Department of Physics, Department of Electrical and
Computer Engineering, and Institute for Research in Electronics and
Applied Physics, University of Maryland, College Park, Maryland
20742, USA }

\author{Brian R. Hunt}
\affiliation{Department of Mathematics and Institute for Physical
Science and Technology, University of Maryland, College Park,
Maryland 20742, USA}

\date{\today}

\begin{abstract}
The largest eigenvalue of the adjacency matrix of a network plays an
important role in several network processes (e.g., synchronization
of oscillators, percolation on directed networks, linear stability
of equilibria of network coupled systems, etc.). In this paper we
develop approximations to the largest eigenvalue of adjacency
matrices and discuss the relationships between these approximations.
Numerical experiments on simulated networks are used to test our
results.

\end{abstract}


\maketitle
\section{Introduction}

In recent years, there has been much interest in the study of the
structure of networks arising from real world systems
\cite{newman1}. Another concern has been dynamical processes taking
place on networks, and the impact of network structure on such
dynamics. The largest eigenvalue of the network adjacency matrix has
emerged as a key quantity important for the study of a variety of
different dynamical network processes. In particular, this is the
case for synchronization of coupled heterogeneous oscillators
\cite{onset}, percolation on directed networks \cite{perco2}, linear
stability of the fixed points of systems of network-coupled ordinary
differential equations \cite{may}, and several other examples in
physics and chemistry \cite{survey,siamlambda}. In this paper we
study methods of obtaining approximations to the maximum eigenvalue
for the adjacency matrix for the case of large complex networks.

We consider a network as a directed graph with $N$ nodes, and we
associate to it an $N\times N$ adjacency matrix whose elements
$A_{ij}$ are one if there is a directed link from $i$ to $j$ and
zero otherwise ($A_{ii} \equiv 0$). We denote the largest eigenvalue
of $A$ by $\lambda$ (it may be shown that of all the eigenvalues of
$A$, the one with largest magnitude is real and positive
\cite{siamlambda}). Furthermore, we note that it is often the case
that the largest eigenvalue is well separated from the second
largest eigenvalue.

The properties of $\lambda$ have been studied in the context of
small or regular graphs \cite{survey} and in classical Erd\"os-Renyi
random graphs \cite{krivelevich}. However, the structure of real
world networks is usually more complex, as evidenced by the fact
that the degree distribution in a large number of examples has been
found to be highly heterogeneous (often following a power law
\cite{barabasi0}), where the {\it out-degree} and {\it in-degree} of
a node $i$ are defined by $d_i^{out}=\sum_{j=1}^N A_{ij}$ and
$d_i^{in}=\sum_{j=1}^N A_{ji}$.
 The `degree distributions', $\hat
P(d^{in},d^{out})$, $P_{in}(d^{in}) = \int
P(d^{in},d^{out})dd^{out}$,
 and $P_{out}(d^{out}) = \int P(d^{in},d^{out})dd^{in}$, are defined as the probabilities
that a randomly chosen node has degree $d^{in}$ and/or $d^{out}$. If
$A =  A^T$ we say the graph is undirected. For an undirected graph
each link serves as both an in and out link for each of the two
nodes it joins and for any node $i$ we have $d_i^{in} =
d_i^{out}\equiv d_i$. Thus in the undirected case we write $P(d)$ to
denote the corresponding degree distribution. The effect of the
degree distribution on the largest eigenvalue of the adjacency
matrix has been explored recently by Chung {\it et al.}
\cite{chung}, who considered a particular ensemble of random {\it
uncorrelated}, undirected networks whose number of nodes $N$ is
large (see also \cite{farkas,goh,mendes}). Here, by uncorrelated we
mean that we regard the network to be a random draw from some
ensemble of networks for which the joint probability distribution of
the node degrees $Q(d_1,d_2,\dots,d_N)$ factors,
$Q(d_1,d_2,\dots,d_N) = P(d_1)P(d_2)\dots P(d_N)$. Chung {\it et
al.} found that in the limit $N\to\infty$ these networks yield an
expected largest eigenvalue that is determined  by the ratio $\hat
\lambda$ of the second to first moment of the average degree
distribution,
\begin{equation}\label{d2d1}
\hat \lambda = \langle d^2 \rangle/\langle d \rangle,
\end{equation}
where $\langle x \rangle = N^{-1}\sum_{i=1}^N x_i$, and by the
expected largest degree $\bar d_{max}$. Specifically, they found
that
\begin{equation}\label{chunga}
\lambda \approx \left\{ \begin{array}{ll}
         \hat \lambda \hspace{1cm}  \hspace{1cm} \hat \lambda > \bar d_{max} \log N,\\
         \sqrt{\bar d_{max}}    \hspace{1.2cm} \sqrt{\bar d_{max}} > \hat \lambda \log^2 N.\end{array} \right.
\end{equation}
Some previous results for dynamical processes in networks have been
stated in terms of the quantity $\hat \lambda$, for example, the
synchronization threshold in the {\it mean field theory} of coupled
oscillators in networks \cite{onset,ichinomiya,lee} and the network
percolation and epidemic spreading thresholds
\cite{cohenperco,moreno0}.

Real world networks often have some amount of degree-degree
correlations \cite{mixing}, i.e., a node of a given degree is more
likely to be connected to nodes with particular degrees than would
be expected on the basis of chance. Networks in which high degree
nodes connect preferentially to high (low) degree nodes, and
viceversa, are called {\it assortative} ({\it disassortative}). Such
correlations can affect dynamical process on networks, as has been
demonstrated for example in epidemic spreading models and
percolation \cite{boguna,newman2,moreno1}.

We also emphasize that the in- and out- degrees at a node can have
different distributions [i.e., $P_{in}(d^{in})\neq
P_{out}(d^{out})$], as has been noted for some corporate information
and genetic networks \cite{braha,guelzim}, and that there are
potential correlations between the in- and out- degrees at the {\it
same} node which can also significantly affect the largest
eigenvalue.

The rest of this paper is organized as follows. Section
\ref{correlations} reviews the characterization of degree
correlations. Section \ref{corre} develops the theory of the maximum
eigenvalue $\lambda$ for the case of networks that satisfy a certain
Markovian property. Some of the considerations of Sec.~\ref{corre}
are similar to theory in previous papers, where, however, those
previous considerations were application specific, and the more
general applicability to the largest eigenvalue was not apparent.
Section \ref{sempling} considers results for numerical constructions
of network adjacency matrices of different types.

\section{Degree-degree correlations}\label{correlations}

\subsection{In and out degree correlations}

As we discussed in the introduction, we define $\hat
P(d^{in},d^{out})$ as the probability that a randomly chosen node
has in-degree $d^{in}$ and out-degree $d^{out}$.

As we shall see later, for networks without neighbor degree-degree
correlations, a first order approximation to the eigenvalue
$\lambda$, generalizing Eq.~(\ref{d2d1}), is given by
\begin{equation}\label{mefi}
\hat \lambda = \langle d^{in} d^{out}\rangle/\langle d \rangle,
\end{equation}
where we recall that $\langle\cdot\rangle$ denotes an average over
nodes, and $\langle d \rangle$ means either $\langle d_i^{in}
\rangle$ or $\langle d_i^{out} \rangle$ which are equal: $\langle
d_i^{in} \rangle = N^{-1}\sum_i d_i^{in} = N^{-1}\sum_{i,j} A_{ij}=
\langle d_i^{out}\rangle$.

If $d_i^{in}$ and $d_i^{out}$ are independent, the largest
eigenvalue is approximately given by $\hat \lambda = \langle d
\rangle$, and if $d_i^{in}$ and $d_i^{out}$ are perfectly
correlated, so that $ d_i^{in} = d_i^{out} \equiv d_i$ (e.g., as in
an undirected network), the largest eigenvalue is approximately
$\hat \lambda = \langle d^2\rangle/\langle d \rangle$. We see that
correlations between $d^{in}$ and $d^{out}$ can crucially affect the
eigenvalue, especially if the second moment of either degree
distribution diverges with increasing $N$, while the first moment
converges: in such a case independence leads to a finite eigenvalue
estimate, and perfect correlation to a diverging eigenvalue
estimate.

In order to quantify the correlations between $d^{in}$ and $d^{out}$
at a node, we define the node degree correlation coefficient,
\begin{equation}\label{eta}
\eta \equiv \langle d^{in} d^{out}\rangle/\langle d \rangle^2.
\end{equation}
Note that if there are no correlations, $\eta = 1$, and $\eta$ is
larger (smaller) than $1$ for positive (negative) correlations. In
terms of the correlation coefficient, $\hat \lambda$ is given by
\begin{equation}\label{etalam}
\hat \lambda = \eta \langle d\rangle.
\end{equation}

\begin{figure}[h]
\begin{center}
\epsfig{file = 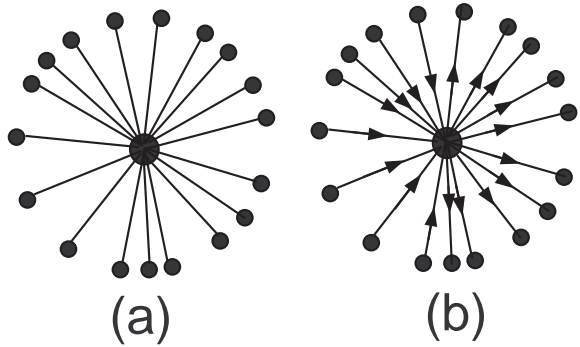, clip =  ,width=1.0\linewidth }
\caption{Undirected (a) and directed (b) star networks illustrating
the fact that, in the directed case, a node with many in and out
links does not necessarily constitute a subnetwork with large
eigenvalue (see text).} \label{nodo}
\end{center}
\end{figure}

For undirected networks $A$ is a symmetric positive matrix, and thus
its largest eigenvalue satisfies
\begin{equation}\label{qq}
\lambda \geq \frac{q^TAq}{q^Tq},
\end{equation}
for any $N$-vector $q$. Choosing the components of $q$ to be zero
for nodes not connected to the node of largest degree $d_{max}$, to
be one for nodes connected to the node of largest degree, and
$\sqrt{d_{max}}$ for the node of largest degree, Eq.~(\ref{qq})
yields
\begin{equation}\label{dmax}
\lambda \geq \sqrt{d_{max}}.
\end{equation}
If $\sqrt{d_{max}} > \hat \lambda$, the mean-field approximation
(\ref{mefi}) must be incorrect. Thus, for undirected networks we use
as a heuristic alternative to the mean-field approximation,
\begin{equation}\label{dmax2}
\lambda \approx \max\{\hat \lambda,\sqrt{d_{max}}\},
\end{equation}
which is consistent with both of the regimes considered in the
rigorous result (\ref{chunga}). (We remark, however, that
Eq.~(\ref{dmax2}) holds in principle only if the ratio $\hat
\lambda/\sqrt{d_{max}}$ or its inverse is large enough [see
Eq.~(\ref{chunga})]). One way of viewing Eq.~(\ref{dmax2}) is that
Eq.~(\ref{qq}) implies that $\lambda$ is at least as large as the
maximum eigenvalue of any subnetwork of the original network (by a
subnetwork we mean one obtained by deleting links of the original
network). Considering a subnetwork consisting of the node of maximum
degree and the nodes connecting to it, shown in Fig.~\ref{nodo} (a),
Eq.~(\ref{dmax}) corresponds to the fact that $\sqrt{d_{max}}$ is
the maximum eigenvalue of this star network. The regime $\lambda
\approx \sqrt{d_{max}}$ in Eq.~(\ref{dmax2}) applies to networks
whose largest eigenvalue is dominated by the node with largest
degree.

We now contrast the above situation for undirected networks with
what can happen for directed networks. We first note that the
reasoning leading to (\ref{dmax2}) may not hold. For example,
Eq.~(\ref{qq}) no longer applies, and a node with many in and out
links does not necessarily make a subnetwork with large eigenvalue.
Regarding the latter point, consider Fig.~\ref{nodo}(b) as compared
to Fig.~\ref{nodo}(a). This directed network has all its eigenvalues
zero, because no pair of in and out links connect the same two
nodes. Moreover, if one views $d^{in}$ and $d^{out}$ to diverge as
$N$ increases at a rate sufficiently more slowly than $N$, then, as
$N$ approaches infinity, the probability of an out link and an in
link from the central node connecting to the same peripheral node
goes to zero with increasing $N$, if these peripheral nodes are
randomly chosen.

\subsection{Neighbor degree correlations}

For our subsequent analysis it is useful to introduce an edge degree
correlation coefficient $\rho$ characterizing the correlation
between the in-degree at node $i$ and the out-degree at node $j$,
where a directed link goes from $i$ to $j$,
\begin{equation}\label{rho}
\rho \equiv \langle d_i^{in} d_j^{out}\rangle_e/\langle
d_i^{in}\rangle_e\langle d_j^{out} \rangle_e,
\end{equation}
where $\langle Q_{ij}\rangle_e$ denotes an average over edges,
\begin{equation}\label{erage}
\langle Q_{ij}\rangle_e \equiv
\sum_{i,j}A_{ij}Q_{ij}/\sum_{i,j}A_{ij}.
\end{equation}
Using (\ref{erage}) we have
\begin{equation}
\langle d_i^{in}\rangle_e =
\sum_{i,j}A_{ij}d_{i}^{in}/\sum_{i,j}A_{ij}=
\sum_{i}d_{i}^{out}d_{i}^{in}/\sum_{i,j}A_{ij}\nonumber
\end{equation}
\begin{equation}\label{ce}
= \langle d^{out}d^{in}\rangle/\langle d \rangle,
\end{equation}
and so
\begin{equation}
\rho = \frac{\langle d_i^{in} d_j^{out}\rangle_e}{\eta^2 \langle d
\rangle^2},
\end{equation}
(Our definition of the edge degree correlation coefficient $\rho$ is
related to but slightly different from that of Newman
\cite{mixing}.)

We note that the term $\sum_{i,j}A_{ij}d_i^{in} d_j^{out}$ that
appears in the definition of $\langle d_i^{in} d_j^{out} \rangle_e$
is the number of directed paths of length three, i.e., the number of
links of the form $n\to i\to j \to m$. Similarly, $\langle
d_i^{in}\rangle_e N \langle d\rangle = \langle d_j^{out} \rangle_e N
\langle d\rangle = N \langle d^{in} d^{out}\rangle =
 \sum_{k}d_k^{out}d_k^{in} =
 \sum_{i,j,k}A_{ik}A_{kj}$
is the number of paths of length $2$, and $N \langle d \rangle$ is
the number of paths of length $1$ (or the number of links).
Accordingly, let us write $\sum_{i,j}A_{ij}d_i^{in} d_j^{out}
\equiv\hspace{0.2cm} n_3$, $N \langle d^{in} d^{out}\rangle
\equiv\hspace{0.2cm} n_2$, and $N \langle d \rangle
\equiv\hspace{0.2cm} n_1$. With this notation, the coefficients
$\eta$ and $\rho$ can be rewritten as
\begin{equation}\label{rflechas}
\eta = \frac{n_2}{n_1 \langle d\rangle},
\end{equation}
\begin{equation}\label{rflechas2}
\rho = \frac{n_3 n_1}{n_2^2},
\end{equation}
As an example of this interpretation, we note that for networks with
uncorrelated in- and out-degrees, the number of paths of length two
is $n_2\approx n_1 \langle d\rangle$, and so $\eta \approx 1$. For
networks where there are no neighbor degree correlations, the number
of paths of length three, $n_3$, can be obtained from the number of
paths of length two, $n_2$, times the average branching ratio given
by $n_2/n_1$, and thus $\rho \approx 1$ for such networks.

\section{Largest eigenvalue of Markovian networks}\label{corre}

\subsection{Formulation}\label{formulation}

For generality, in this section we allow different types of network
nodes, where we specify the node type by an index $\sigma =
1,2,\dots,M$, where $M$ is the total number of possible node types
(e.g., in the case of social networks connecting people, $\sigma$
might label sex, race, social class, etc.). Furthermore, we
introduce the quantity
\begin{equation}
z = (d^{in},d^{out},\sigma),
\end{equation}
which we refer to as the degree. With this definition, we use $P(z)$
to denote the degree distribution, i.e., $P(z)$ is the probability
that a randomly chosen node has degree $z$. This implies, for
example, that
\begin{equation}
\sum_{d^{in},d^{out}}P(z)=N_{\sigma}/N,
\end{equation}
where $N_\sigma$ is the number of nodes of type $\sigma$ and $N =
\sum_\sigma N_\sigma$. (While our numerical examples in
Sec.~\ref{sempling} are for the case of a single node type, $M=1$,
the subsequent considerations in the present section do not have
this restriction.)

We consider a particular class of networks for which the only
nontrivial correlations are between nodes that are directly
connected by a single link. Such {\it Markovian} networks have been
considered in previous works on epidemic spreading and percolation
\cite{cohenperco,moreno0,moreno1,boguna}. Under this assumption, if
we define $P(z'|z)$ to be the probability that a node with degree
$z$ has an outgoing link to a node with degree $z'$, then if we
choose a random outward path of length two from a node with degree
$z$, the probability that the first hop ends on a node with degree
$z'$ and the second on a node with degree $z''$ is, by this
assumption,
$$
P(z',z''|z) = P(z''|z')P(z'|z).
$$

Let $\psi_z^{(m)}$ be the expected number of directed paths of
length $m$ whose starting node has degree $z$. Using the assumption
that the network is Markovian, we can express $\psi^{(m+1)}_z$ in
terms of $\psi^{(m)}_z$ as
\begin{equation}\label{psis}
\psi^{(m + 1)}_z = d^{out}\sum_{z'}P(z'|z)\psi_{z'}^{(m)}.
\end{equation}
The number of paths of length $m$ grows, in the limit of large $m$,
as the largest eigenvalue $\lambda$, $\psi_{z}^{(m)}\sim \psi_z
\lambda^m $ \cite{survey}. Therefore, we can associate to
Eq.~(\ref{psis}) the eigenvalue problem
\begin{equation}\label{evp}
\lambda_C \psi_z = d^{out}\sum_{z'}P(z'|z)\psi_{z'},
\end{equation}
and we seek its largest eigenvalue $\lambda_C$ and its corresponding
eigenfunction $\psi_z$. We consider $\lambda_C$ to be an
approximation to $\lambda$ that is more accurate than the mean field
result in that it includes correlations between connected nodes,
$\lambda \approx \lambda_C$. Additionally, an approximation $u_C$ to
the right eigenvector $u$ of $A$ can be obtained in terms of
$\psi_z$ in a similar way:
\begin{equation}\label{uc}
(u_C)_i = \psi_{z_i},
\end{equation}
and an analogous expression holds for the left eigenvector $v$ of
$A$, using the left eigenvector of the matrix $d^{out}P(z'|z)$
instead of $\psi_z$. We will refer to Eqs.~(\ref{evp}) and
(\ref{uc}) as the {\it Markovian approximation}.

\subsection{$\epsilon$ expansion}\label{expansion}

While in many cases Eq.~(\ref{evp}) can be solved directly, it is
also of interest to explore approximations to its solution. Thus, we
will here expand (\ref{evp}) about a zero order approximation in
which degree-degree correlations are neglected and in which the
first order correction gives the perturbation to the uncorrelated
case due to a small amount of degree-degree correlations. In
particular, we will be interested in zeroth order and first order
approximations to the largest eigenvalue of $A$ and to the
corresponding right and left eigenvectors $u$ and $v$, where
\begin{equation}
Au=\lambda u \hspace{2mm} \mbox{    and    } \hspace{2mm} v^T A =
\lambda v^T.
\end{equation}
Following previously used terminology, we refer to the zeroth
approximation as the {\it mean field theory}.

When there are no neighbor degree-degree correlations, $P(z'|z)$
becomes independent of $z$:
\begin{equation}\label{uncorre}
P(z'|z) = \hat P(z') = (d^{in})' P(z')/\langle d \rangle,
\end{equation}
where $P(z)$ is the degree distribution. We will expand
Eq.~(\ref{evp}) about the uncorrelated case. For this purpose, we
will write
\begin{equation}
P(z'|z) = \hat P(z') + \epsilon \delta P(z'|z),
\end{equation}
where
\begin{equation}
\delta P(z'|z) = P(z'|z) - \hat P(z'),
\end{equation}
and $\epsilon$ is an expansion parameter that we formally consider
small, although, in reality, $\epsilon = 1$. Introducing expansions
for $\lambda_C$ and $\psi_z$,
\begin{equation}
\lambda_C = \hat \lambda + \epsilon \delta \lambda +\dots
\end{equation}
\begin{equation}
\psi_z = \hat \psi_z + \epsilon \delta \psi_z +\dots
\end{equation}
Expanding Eq.~(\ref{evp}) to zero order in $\epsilon$, we obtain
\begin{equation}
\hat \psi_z = d^{out},
\end{equation}
and
\begin{equation}\label{mefe}
\hat \lambda = \langle d^{in}d^{out}\rangle/ \langle d\rangle.
\end{equation}
Thus in the zeroth order approximation, the right eigenvector of $A$
has components
\begin{equation}
u_i = d_i^{out}.
\end{equation}
To obtain the left eigenvector, we follow the same steps as above
but with $A$ replaced by $A^T$. This interchanges the roles of
$d^{in}$ and $d^{out}$, thus yielding
\begin{equation}
v_i = d_i^{in}
\end{equation}
for the left eigenvector.

Expanding to first order in $\epsilon$, we obtain
\begin{equation}
\hat \lambda \delta \psi_z + \delta \lambda \hat \psi_z = d^{out}
\sum_{z'} \hat P(z') \delta \psi_{z'} + d^{out} \sum_{z'} \delta
P(z'|z)\hat \psi_{z'}.
\end{equation}
Multiplying by $\hat P(z)$ and summing over $z$ we obtain, after
some simplification,
\begin{equation}
\delta \lambda = \left[(\hat \lambda \langle d\rangle)^{-1}
\sum_{z,z'} (d^{out})' d^{out} d^{in}P(z'|z)P(z)\right] - \hat
\lambda
\end{equation}
The probability that a randomly chosen edge starts at a node with
degree $z$ is $\tilde P(z) = \langle d\rangle^{-1} P(z) d^{out}$,
and, therefore, the term $\langle d\rangle^{-1} \sum_{z,z'}
(d^{out})' d^{out} d^{in}P(z'|z)P(z)$ is equal to
\begin{equation}
\sum_{z,z'} d^{in} (d^{out})'\tilde P(z') P(z'|z)= \langle
d_i^{in}d^{out}_j\rangle_e.
\end{equation}
To first order, therefore, we obtain
\begin{equation}\label{linear}
\lambda_C \approx \hat \lambda  + \delta \lambda = \frac{\langle
d_i^{in}d^{out}_j\rangle_e}{\hat \lambda} = \hat \lambda \rho,
\end{equation}
where $\rho$ is defined in Sec.~\ref{correlations}, and we call
$\lambda = \hat \lambda \rho$ the {\it linear approximation}.

Similarly,
\begin{equation}
\delta \psi_z = \hat \lambda^{-1}\sum_{z'} \delta P(z'|z)
(d^{out})'.
\end{equation}
\begin{figure}[h]
\begin{center}
\epsfig{file = 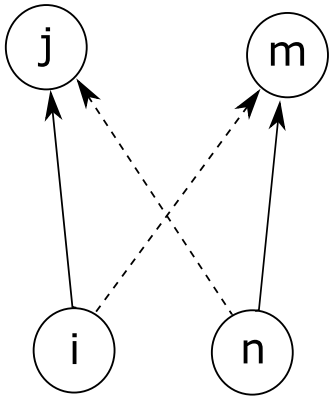, clip =  ,width=0.5\linewidth }
\caption{Schematic representation of the rewiring algorithm. Edges
$i \to j$ and $n \to m$ were chosen at random. When creating a
network with assortative (disassortative) correlations, they are
replaced with edges $i \to m$ and $n \to j$ if the resulting network
has a larger (smaller) value of $\rho$.} \label{rewire}
\end{center}
\end{figure}

We now briefly comment on the range of validity of the expansion
results. As the rigorous result Eq.~(\ref{chunga}) shows, the mean
field result for undirected networks is not valid when the network
is dominated by the node with maximum degree. A network in which the
maximum degree is too large compared with the bulk of the degree
distribution can not satisfy Eq.~(\ref{uncorre}). For example, the
star network of Fig.~\ref{nodo}(a) is very degree-degree correlated
because all of its outer nodes connect only to the high degree hub
node and, indeed, application of Eq.~(\ref{uncorre}) to this simple
star network yields an incorrect result. Our expansions about an
uncorrelated network therefore implicitly assumed that this network
was not dominated by the node with maximum degree. We speculate that
directed networks might satisfy Eq.~(\ref{mefe}) with a less
restrictive condition on the maximum degrees, as argued in
Sec.~\ref{correlations} and Fig.~\ref{nodo}. A more rigorous
delineation of the range of validity of the Markovian approximation
for correlated and/or directed networks along the lines of
Ref.~\cite{chung} is open for further research.

\section{Numerical tests on simulated networks}\label{sempling}

In this section we numerically construct random networks and use
them to compare with the predictions of Sec.~\ref{corre}. We shall
restrict ourselves to networks which are not dominated by nodes with
large degree, as discussed at the end of Sec.~\ref{expansion}. In
order to study the variation of the largest eigenvalue $\lambda$ of
the adjacency matrix $A$, we first construct large ($N\gg 1$)
approximately uncorrelated networks with a given expected degree
distribution using a generalization of
 the Random Graph model of
Chung {\it et al.} \cite{chung}. First we associate to each node $k$
target in and out degrees $(\hat d_k^{in},\hat d_k^{out})$, where
$\sum_k \hat d_k^{in} = \sum_k \hat d_k^{out} = N \langle \hat
d\rangle$. Note that the choice of target node degrees can be made
to correspond to a desired degree distribution (e.g., scale free, or
Poisson). We then choose each element of the adjacency matrix
$A_{ij}$ randomly to be one with probability $\hat d_i^{out} \hat
d_j^{in}/(N\langle \hat d\rangle)$ and zero otherwise. This
determines a network realization with in and out degrees $d_i^{out}
= \sum_j A_{ij}$ and $d_j^{in} = \sum_i A_{ij}$. In general,
$(d_k^{in},d_k^{out})$ can be different from the target values
$(\hat d_k^{in},\hat d_k^{out})$. Nevertheless, with high
probability, for large $N$, the resulting degree distribution
$P(d^{in},d^{out})$ of this randomly chosen network will be
approximately, in a suitable sense, the target distribution
\cite{chung}. In particular, the moments $\langle d^{in}
d^{out}\rangle$ and $\langle d\rangle$ will be approximately
unchanged when calculated using either $P$ or the target degrees.
Furthermore, $\rho\approx 1$ for $N\gg 1$. The network generated by
this algorithm is directed (i.e., $A$ is asymmetric); symmetric
networks can be generated by first considering only $i < j$ and then
setting $A_{ji} = A_{ij}$.
\begin{figure}[h]
\begin{center}
\epsfig{file = 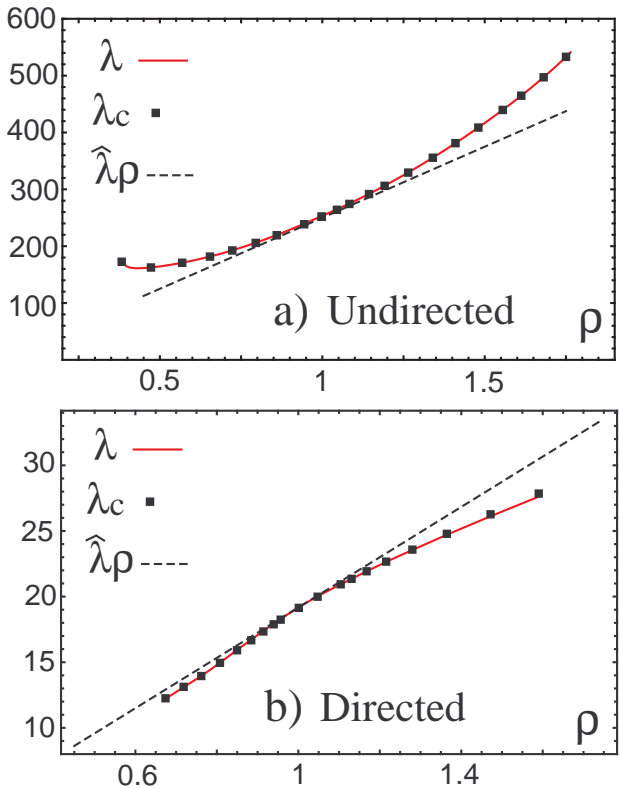, clip =  ,width=1.0\linewidth } \caption{
Largest eigenvalue $\lambda$ (solid line), linear approximation
[Eq.~(\ref{linear}), dashed line], and Markovian approximation
$\lambda_C$ [Eq.~(\ref{evp}), boxes] for a) an undirected network
with $N= 25000$, $\gamma = 2.5$ and $\langle d\rangle = 100$, and b)
a directed network with $N= 10000$, $\gamma = 2.5$ and $\langle
d\rangle = 20$.} \label{fig1}
\end{center}
\end{figure}
Starting from an uncorrelated network generated by this algorithm,
we then rewire the connections in such a way that the degree
distribution is preserved. On doing so, the correlation coefficient
changes and we calculate the largest eigenvalue for different values
of $\rho$. The rewiring algorithm we use is a simplified version of
that used in \cite{newman2} and consists of the repeated iteration
of the following steps:
\begin{enumerate}

\item Two edges are chosen at random. Assume one connects node
$n$ to node $m$ and the other connects node $i$ to node $j$.
\item Let
\begin{equation}\label{t}
H(i,j;n,m) = d_n^{in}d^{out}_m +
d_i^{in}d^{out}_j-d_n^{in}d^{out}_j-d_i^{in}d^{out}_m.
\end{equation}
The two edges chosen in step $1$ are replaced with two edges
connecting node $n$ to node $j$ and node $i$ to node $m$ (see
Fig.~\ref{rewire})  if $s H(i,j;n,m) < 0$, and are left alone
otherwise.  Setting $s = 1$ or $-1$ we produce assortative or
disassortative networks, respectively.
\end{enumerate}

As we iterate from $\rho=1$ with $s = 1$ $(s = -1)$, $\rho$ steadily
increases (decreases). Thus, we produce a sequence of networks with
successively larger (smaller) $\rho$.

\begin{figure}[t]
\begin{center}
\epsfig{file = 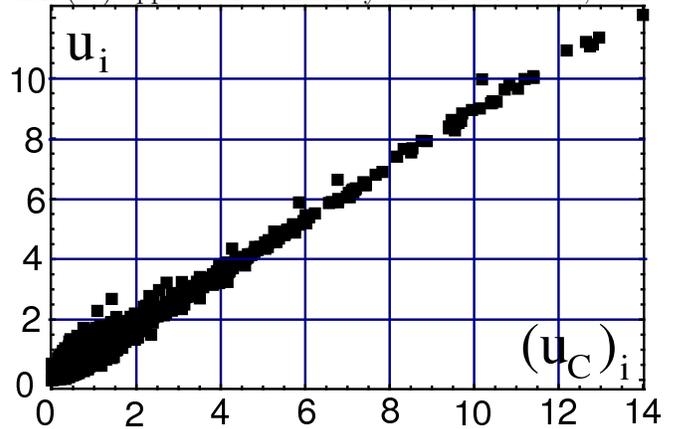, clip =  ,width=1.0\linewidth } \caption{
Entries of the eigenvector $u$ corresponding to $\lambda$ versus the
entries of its Markovian approximation $u_C$ [Eq.~(\ref{uc})] in an
arbitrary scale for the network of Fig.~\ref{fig1} b) at $\rho
\approx 0.9$. } \label{figg2}
\end{center}
\end{figure}

In Fig.~\ref{fig1} a) we show the largest eigenvalue $\lambda$
(solid line), the linear approximation given by Eq.~(\ref{linear})
(dashed line), and the Markovian approximation $\lambda_C$ from
Eq.~(\ref{evp}) (boxes) as a function of the correlation coefficient
$\rho$ for an undirected network with power law degree distribution
generated as described above with $N = 25000$, $\langle d \rangle =
100$, and exponent $\gamma = 2.5$. In Fig.~\ref{fig1} b) we plot the
same quantities for a directed network with $N = 10000$, $\langle d
\rangle = 20$, and exponent $\gamma = 2.5$. In these plots there is
no discernible difference between the approximation $\lambda_C$ and
the actual values of $\lambda$. We observe that the largest
eigenvalue depends strongly on the correlation coefficient: in the
undirected case, it increases more than three times as $\rho$ varies
from $0.4$ to $1.7$. Also, we see that in these examples the linear
approximation works for $|\rho-1| \lesssim 0.2$, but fails for
larger values of $|\rho-1|$. In the undirected case, $\lambda$ is
larger than the linear approximation, which follows from
Eq.~(\ref{qq}) if we set $q_i = d_i$.

In Fig.~\ref{figg2} we show the eigenvector $u_i$ for the network of
Fig.~\ref{fig1} b) at $\rho \approx 0.9$ plotted against the
corresponding approximation $(u_C)_i$, Eq.~(\ref{uc}) using an
arbitrary scale. There is good agreement between the true value and
its Markov estimate.

\section{Conclusion}\label{discuss}

In this paper we have considered several approximations to the
largest eigenvalue of the adjacency matrix of large, directed
networks. The mean field result (\ref{mefi}) appears to apply well
to networks whose neighboring nodes are uncorrelated in their
degrees. The linear approximation (\ref{linear}) applies for
sufficiently small correlation, while the Markov model (\ref{evp})
applies for arbitrarily strong degree correlations between
neighbors. The price to be paid for a more refined approximation is
the requirement of greater knowledge of the network (e.g., use of
(\ref{evp}) requires knowledge of $P(z|z')$ which is not required
for the two other less refined approximations).

We caution that, although we have obtained good agreement between
the theory and numerical results on simulated networks, this may not
necessarily carry through for real networks encountered in practice.
In particular, the Markov assumption of Eq.~(\ref{psis}) may not
always hold (e.g., due to community structure \cite{girvan},
clustering, or degree correlations extending over more than one link
between nodes). This remains a topic for further study.

This work was supported by ONR (Physics), by the NSF (PHY 0456240
and DMS 0104-087), and by AFOSR FA95500410319.

\end{document}